\documentstyle[12pt]{article}

\newcommand{\beq}{\begin{equation}}
\newcommand{\eeq}{\end{equation}}
\newcommand{\bea}{\begin{eqnarray}}
\newcommand{\eea}{\end{eqnarray}}

\newcommand{\Ds}{\not\!\! D}
\newcommand{\as}{\not\!\! A}
\newcommand{\aas}{/\kern-.50em A}
\newcommand{\qa}{\not\! a}
\newcommand{\ds} {\!\!\not\!\partial}

\newcommand{\unc}{{\bf 1}_c}
\newcommand{\unf}{{\bf 1}_f}

\newcommand{\al}{\alpha}
\newcommand{\mb}{\makebox[0.7cm] }
\newcommand{\D}{{\cal{D}}}

\newcommand{\M}{{\cal M}}

\begin{document}

%\draft

%\preprint{\vbox{ {\it to appear in Physical Review D55} \hfill
%{\tt CBPF-NF-050/96 and La Plata-TH-013/96}\\}}

\title
{Multiflavor Correlation Functions in non-Abelian
Gauge Theories at Finite Fermion Density in two dimensions}
\author{H. R. Christiansen \thanks{On leave from Universidad
Nacional de La Plata and CONICET, Argentina.}\\
{\normalsize\it Centro Brasileiro de Pesquisas Fisicas .
CBPF - DCP}\\
{\normalsize\it Rua Xavier Sigaud 150, 22290-180
Rio de Janeiro, Brazil.}\\
{}~and\\
F. A. Schaposnik\thanks{Investigador CIC.}\\
{\normalsize\it Departamento de F\'\i sica, Universidad
Nacional de La Plata}\\ {\normalsize\it C.C. 67, (1900) La Plata,
Argentina.}}

\maketitle

\def\thepage{\protect
\vbox{{\it to appear in Physical Review D55}
\hfill{Last version}\\}}
\thispagestyle{headings}
%\markright{\thepage}

\begin{abstract}
\normalsize
We compute vacuum expectation values of products of fermion
bilinears for two-dimensional Quantum Chromodynamics at finite
flavored fermion densities. We  introduce the chemical
potential as an external charge distribution
within the path-integral approach and carefully
analyse the contribution of different topological sectors
to fer\-mion correlators. We show the existence of chiral
condensates exhibiting an oscillatory inhomogeneous behavior
as a function of a chemical potential matrix. This result is
exact and goes in the same direction as the behavior found
in {\em QCD}$_4$ within the large $N$ approximation.
\end{abstract}


\newpage
\pagenumbering{arabic}
\section{Introduction \label{sec-intro}}

In order to  understand the structure of the {\em QCD}
vacuum \cite{shuryak}
one should analyse possible mechanisms for chiral symmetry breaking
and the formation of fermion condensates.
The existence of such correlators can be
understood as the result of
condensation of pairs of particles and holes
and it can have interesting implicancies 
in particle physics and cosmology. For example, a color nonsinglet
condensate may be related to superfluidity and color superconductivity
of cold quark matter at high fermion densities \cite{bl}.
In this respect the results of Deryagin, Grigoriev and Rubakov \cite{dgr}
are of particular importance. Analysing the large $N_c$ limit
of ${\em QCD}$ these authors have shown that the order parameter for
chiral symmetry, the quark condensate $\langle\bar \psi \psi\rangle$,
is at high quark densities inhomogeneous and anisotropic so that,
regarding the order parameter, the ground state of quark matter has
the structure of a standing wave.

Two-dimensional models like the Schwinger model and ${\em QCD}_2$
provide a natural laboratory to test these phenomena since,
although simplified, the basic aspects (chiral symmetry
features, non-trivial topological sectors, etc) are
still present and exact calculations can be in many cases performed.

An analysis of two-dimensional $QED$ at
finite density was originally presented in \cite{fks}-\cite{ara}.
More recently,  studies on this theory \cite{Kao}-\cite{hf}
showed that inhomogeneous chiral condensates do exist as a result
of the contribution of non-trivial topological sectors.

Extending our work on $QED_2$ \cite{hf}
we analyse in the present paper vacuum expectation values
of products of local bilinears $\bar \psi(x) \psi(x)$, at finite
density for two-dimensional Quantum Chromodynamics with flavor.
Using a path-integral approach which is very appropriate
to handle non-Abelian gauge theories, we show that the
multipoint chiral condensates exhibit an oscillatory inhomogenous
behavior depending on a chemical potential matrix.
Our results are exact and, remarkably, go in the same direction
as those revealed in four dimensions using the $1/N_c$ approximation 
to ${\em QCD}$ \cite{dgr}.

To study the effect of finite fermion density in ${\em QCD}_2$
a chemical potential may be introduced. Within the path-integral
approach this ammounts to consider a classical background
charge distribution in addition to that produced by
topologically non-trivial gauge configurations.
Concerning this last point,  it is well-known that in two 
space-time dimensions the role of 
instantons is played by vortices. In the Abelian case, these vortices
are identified with the Nielsen-Olesen solutions
of the spontaneously broken Abelian Higgs model \cite{NO}.
Also in the non-Abelian case,
regular solutions with topological charge exist
when symmetry breaking is appropriately achieved via Higgs fields
\cite{dVS}-\cite{LMS}.
In both cases the associated fermion zero modes have been 
found \cite{JR}-\cite{CL}.

Properties of the vortex solutions and the corresponding
Dirac equation zero-modes are summarized in section 2. We then
describe in sections 3 and 4 how topological effects can be
taken into account within the path-integral formulation
leading to a compact form for the partition function in the
presence of a chemical potential. Our approach,
following ref.
\cite{bc}, starts by decomposing a given gauge field belonging to
the $n^{th}$ topological sector in the form
\beq
A_\mu(x) = A_\mu^{(n)} + A_\mu^{ext} + a_\mu
\label{pi}
\eeq
Here $A_\mu^{(n)}$ is a (classical) fixed gauge
field configuration belonging
to the $n^{th}$ class, $ A_\mu^{ext}$ is the background
charge field taking account of the chemical potential, and $a_\mu$
is the path-integral variable which represents
quantum fluctuations. Both $ A_\mu^{ext}$  and $a_\mu$ belong
to the trivial topological sector and can be then decoupled
by a chiral rotation with the sole evaluation of a Fujikawa jacobian
\cite{Fuj}. This last calculation can be easily performed
since it is to be done in the trivial topological sector.

The complete calculation leading to the minimal non-trivial
correlation functions of fermion bilinears is first presented
for multiflavour $QED_2$ (Section 3) and then extended to multiflavour
$QCD_2$ (Section 4).
In both cases the oscillatory behavior of
correlators as a function of the chemical
potential is computed, the result showing a striking resemblance
with the $QCD_4$ answer obtained within the large $N_c$ approximation
\cite{dgr}. We summarize our results and conclusions
in section 5.

%%%%%%%%%%%%%%%%%%%%%%%%%%%%%%%%%%%%%%%%%%%%%%%%%%%%%%%%%%%%%%%%%%%

\section{Zero Modes}

Topological gauge field configurations and
the corresponding zero-modes of the Dirac equation play a central
role in calculations involving
fermion composites. We sumarize
in this section the main properties of vortices, the relevant topological
objects in the model we shall consider,
both  for the Abelian and non-Abelian cases. We also
present the corresponding Dirac operator zero-modes.

\subsection{The Abelian case}

In two-dimensional Euclidean space-time, topologically  non-trivial
configurations are available since Nielsen and
Olesen \cite{NO}
presented their static $z$-independent vortex. In the $U(1)$
case the topological charge for such a configuration,
working in an arbitrary compact surface
(like a sphere or a torus) is defined as
\beq
\frac{1}{4\pi}\int d^2x \, \epsilon_{\mu \nu} \,
F_{\mu \nu}^{(n)} =  n \in Z
\label{q}
\eeq

A representative gauge field configuration
carrying topological charge $n$ can be written as
\beq
A_\mu^{(n)} = n\, \epsilon_{\mu \nu} \frac{x_\nu}{\vert x \vert}
A(\vert x \vert)
\label{v}
\eeq
with $A (\vert x \vert)$ a function which can be calculated
numerically
(an exact solution exists under certain conditions on coupling constants,
\cite{dVS}). The adequate  boundary conditions are

\beq
A(0) = 0 ~~~,~~~
\lim_{\vert x \vert \to \infty} A(\vert x \vert) = -1
\label{c}
\eeq
There are $\vert n \vert$ zero-modes associated with the Dirac operator
in the background of an $A_\mu^{(n)}$ configuration in a suitable
compactified space-time \cite{bc}. (For the non-compact case see
\cite{JR}). For $n>0$ ($n<0$) they correspond
to right-handed (left-handed) solutions $\eta_R$ ($\eta_L$)
which in terms of light-cone
variables $ z = x_0 + i x_1$ and $ \bar z = x_0 - i x_1$
can be written in the form
\beq
\eta_R^m = \left(\begin{array}{c} z^m h(z,\bar z) \\ 0
\end{array} \right)
\eeq
\beq
\eta_L^m = \left(\begin{array}{c} 0 \\{\bar z}^{-m} h^{-1}(z,\bar z)
\end{array} \right)
\eeq
where $m = 0,1, \ldots , \vert n \vert -1$,
\beq
h(z,\bar z) = \exp[\phi^{(n)}(\vert z\vert )]
\label{f}
\eeq
and
\beq
\frac{d}{d\vert z\vert }\phi^{(n)}(\vert z \vert) = n A (\vert z \vert).
\label{ff}
\eeq
%
%%%%%%%%%%%%%%%%%%%%%%%%%%%%%%%%%%%%%%%%%%%%%%%%%%
\subsection{The non-Abelian case}

As in the Abelian case, two-dimensional gauge field configurations
$A_\mu^{(n)}$ carrying a topological charge $n \in Z_N$ can be found
for the $SU(N)$ case. As explained in ref.\cite{dVS2}
the relevant homotopy group is in this case
$Z_N$ and not $Z$ as in the $U(1)$ case.

Calling $\varphi$ the angle characterizing the direction at infinity, a
mapping $g_n(\varphi) \in SU(N)$ belonging to the $n^{th}$ homotopy
class ($n = 0,1, \ldots, N-1$) satisfies, when one turns around a
close contour,
\beq
g_n(2\pi) = \exp(\frac{2\pi i n}{N}) g_n(0)
\label{su1}
\eeq
Such a behavior can be achieved just by taking $g_n$ in the
Cartan subgroup of the gauge group. For example, in the $SU(2)$ case
one can take
\beq
g_n(\varphi) = \exp [\frac{i}{2} \sigma^3 \Omega_n(\varphi)]
\label{222}
\eeq
with
\beq
\Omega_n(2\pi) - \Omega_n(0) = 2\pi (2 k + n)
\label{2c}
\eeq
Here $n=0,1$ labels the topological charge and $k \in Z$ is a second
integer which conects the topological charge with the vortex
magnetic flux (Only for abelian vortices both quantities coincide).

We can then write a gauge
field configuration belonging to the $n^{th}$ topological
sector in the form
\beq
A_\mu^{(n)} = i A(\vert x\vert )\ g_n^{-1} \partial_\mu g_n
\label{gf}
\eeq
with the boundary conditions
\beq
A(0) = 0 ~~~,~~~
\lim_{\vert x \vert \to \infty} A(\vert x \vert) = -1
\label{cna}
\eeq
These and more general vortex configurations have been
thoroughfully studied in \cite{LMS}-\cite{dVS2}.

Concerning zero-modes of the Dirac operator in the background of
non-Abelian vortices, they have been analysed in refs.\cite{dV}-\cite{CL}.
The outcome is that for topological charge $n>0$ ($n<0$) there are $Nn$
($N\vert n \vert$)
square-integrable zero modes $\eta_L$ ($\eta_R$) analogous to those
arising in the Abelian case. Indeed, one has
\beq
\eta_R^{(m,i) j} = \left(\begin{array}{c} z^m h_{ij}(z,\bar z) \\ 0
\end{array} \right)
\label{naz}
\eeq
\beq
\eta_L^{(m,i) j} =
\left(\begin{array}{c} 0 \\{\bar z}^{-m} h_{ij}^{-1}(z,\bar z)
\end{array} \right)
\label{naz2}
\eeq
with
\beq
h(z,\bar z) = \exp[\phi^{(n)}(\vert z\vert) M]
\label{cui}
\eeq
and
\beq
M = \frac{1}{N} {\rm diag} (1,1, \ldots, 1-N)
\label{M}
\eeq
Here $i,j= 1,2, \ldots, N$ and $m = 0,1, \ldots, \vert n \vert - 1$.
The pair $(m,i)$ labels the $N\vert n \vert$ different zero modes while
$j$ corresponds to a color index.
Due to the ansatz discussed in refs.\cite{LMS}-\cite{dVS2}
for the non-Abelian vortex, the function $\phi^{(n)}(\vert z\vert)$
appearing in eq.(\ref{cui})
coincides with that arising in eqs.(\ref{f})-(\ref{ff}) for the abelian
vortex.

As it happens in the abelian case, the partition function of two
dimensional Quantum Chromodynamics
only picks a contribution from the trivial sector because
$\det(\Ds[A^{(n)}])=0$ for $n\neq 0$ (see eq.(\ref{z1}) below).
In contrast, various correlation functions become non-trivial precisely
for $n\neq 0$ thanks to the ``absortion'' of zero-mode contributions when
Grassman integration is performed.

It is our aim to see how these non-trivial correlators are modified
when a fermion finite density constraint is introduced,
comparing the results with those of the unconstrained
(zero chemical potential) case. As explained in the introduction,
we are motivated by the results of  Deryagin,  Grigoriev
and Rubakov \cite{dgr} in four dimensional $QCD$. They were able
to show,
in the large $N_c$  and high fermion density limits, the existence of
oscillatory condensates (the frequency given by the
chemical potential) which are spatially  inhomogeneous.
For $QED_2$ the same oscillatory behavior was found approximately
in \cite{Kao} and confirmed analytically in
\cite{hf}, by examining an arbitrary number of fermion bilinears
for which  the exact $\mu$-dependence of
fermionic correlators was computed. In order to improve our 
understanding of the
large $N_c$ results found in $QCD_4$, we shall extend in what follows
our two-dimensional approach to the non-Abelian case but before,
we shall consider the case of
flavored $QED_2$ as a clarifying step towards multiflavor $QCD_2$.

%%%%%%%%%%%%%%%%%%%%%%%%%%%%%%%%%%%%%%%%%%%%%%%%%%%%%%%%%%%%%%%%%%%%%

\section{Multiflavour $QED_2$}

We developed in ref.\cite{hf} a path-integral method to compute
fermion composites for Abelian gauge theories including
chemical potential effects. In this section we  briefly describe
our approach while extending our treatment so as to include flavour. 
We then leave for section 4 the analysis of the non-Abelian 
multiflavour $QCD_2$ model at finite density.

\subsection*{(i) Handling the chemical potential in the Abelian case}

We start from the Lagrangian
\beq
L= -\frac{1}{4e^2} F_{\mu\nu} F_{\mu\nu}+
\bar\psi (i\ds +\as -i\M\gamma_0)\psi
\label{abef}
\eeq
\noindent
where  $\psi$ is the fermion field isospinor. A
chemical potential term has been included by considering
the diagonal matrix ${\cal M}$ defined as
\beq
{\cal M}= \mb{diag}( \mu_1\dots \mu_{N_f})
\label{Mu}
\eeq
where $N_f $ is the total number of flavors and $\mu_{k}$ are
Lagrange multipliers carrying a flavour index,
so that each $k$-fermion number is independently conserved.
The corresponding partition function is defined as
\begin{equation}
 Z[\mu_1\dots \mu_{N_f}] =  \int \D\bar\psi \D \psi \D A_\mu
\exp (- \int d^2x\ L).
\label{par}
\end{equation}

Since our interest is the computation of fermionic correlators,
we have to carefully
treat non-trivial topological configurations of the gauge fields
which have been seen to be crucial in the obtention of
non-vanishing condensates, see refs.\cite{mnt}-\cite{cmst}.
Then, following the approach of refs.\cite{bc}-\cite{cmst}, we
decompose gauge field configurations belonging to the $n^{th}$
topological sector in the form
\beq
A_{\mu}(x) = { A}_{\mu}^{(n)}(x) + a_{\mu}(x)
\eeq
where $A_{\mu}^{(n)}$ is a fixed classical configuration carrying all
the topological charge $n$, and $a_{\mu}$, the path integral variable,
accounts for the quantum ``fluctuations'' and belongs to the trivial
sector $n=0$.

As it is well-known \cite{Actor}, the chemical potential term can be
represented by a vector field $A_\mu^{ext}$  describing
an {\em external} charge density acting on the quantum system.
Indeed, taking  $A_\mu^{ext}$  as $i$ times the chemical potential
matrix (see eqs.(\ref{Mu}) and (\ref{achem})) it corresponds
to a uniform charge background for each fermionic flavor.
As explained in \cite{hf}, it is convenient
to  first consider a finite length ($2 l$)
box and then take the $l\rightarrow\infty$ limit. In this
way translation symmetry breaking associated to the
chemical potential becomes apparent and simultaneously,
ambiguities in the definition of the finite density theory
are avoided. When necessary, we shall follow this prescription 
(see ref.\cite{fks} for a discussion on this issue).
We start by defining
\beq
A_{\nu}^{ext}=-i{\cal M}\ \delta_{\nu 0},
\label{achem}
\eeq
so that the  Dirac operator
\beq
i\ds+\as-i{\cal M}\gamma_0
\eeq
can be compactly written as
\beq
i\ds +\as'
\label{a'}
\eeq
with
\beq
A'_\mu = A_\mu + A_{\mu}^{ext}
\label{bol}
\eeq

We shall now proceed to a decoupling of fermions from the chemical 
potential and the $a_\mu$ fluctuations following the steps
described in \cite{hf} for the case of only one flavor.
In that case,  we wrote
\beq
a_\mu = -\epsilon_{\mu \nu} \partial_\nu \phi+\partial_{\mu}\eta
\label{viejo}
\eeq
and made a chiral rotation to decouple both the $\phi-\eta$ fields
together with the chemical potential. In order to include $N_f$
flavors in the analysis,  one has
to replace $(\phi,\eta) \rightarrow (\phi,\eta)\unf$ and
$\mu \rightarrow  {\cal M}$ as we shall see below.
Then, we can straightforwardly apply what we have learnt for one
flavor \cite{hf} in the multiflavor case.
The   change of variables accounting for the decoupling
of fermions from the $a_{\mu}$ field together with the
chemical potential is given by
\begin{eqnarray}
& & \psi = \exp[\gamma_5\ (\phi(x)\unf+i\M x_1)+i\eta(x) \unf]\
\chi \nonumber\\
\label{change}\\
& & \bar\psi =  \bar\chi\ \exp[{ \gamma_5\ (\phi(x)\unf+i\M x_1)
-i\eta(x)\unf}]\nonumber
\end{eqnarray}
\beq
\qa=(-i\ds U)\ U^{-1}
\label{decou}
\eeq
where
\beq
U=\exp[{\gamma_5\ (\phi\unf+i\M x_1)+i\eta\unf}]
\eeq
For notation compacteness we have included in
$ \not\!\! a $ the external field $A^{ext}_{\mu}$ describing the chemical
potential term.  From here on we choose the Lorentz gauge to work in
(which in our notation corresponds to  $\eta=0$).

After transformation (\ref{change})  the resulting  Dirac operator
takes the form
\beq
i\Ds=i\ds+\as^{(n)}+\qa\ \  \rightarrow\ \  i\ds+\as^{(n)}.
\eeq
The jacobian associated with the chiral rotation of
the fermion variables can be easily seen to be \cite{hf}
\begin{equation}
J  =  \exp \left(\frac{tr_f}{2\pi}\int d^2x ~(\phi\unf+i\M x_1)
\Box (\phi + 2 \phi^{(n)})
\right)
\end{equation}
where $\phi^{(n)}$ is defined by
\[
A_\mu^{(n)} = - \epsilon_{\mu \nu} \partial_\nu \phi^{(n)} \unf
\]
Together with eq.(\ref{change}) we consider the change in the
gauge-field variables $a_\mu$ so that
\beq
\D a_\mu = \Delta_{FP}\delta(\eta) \D\phi \D\eta
\label{ca}
\eeq
with $ \Delta_{FP} = \det \Box$

As thoroughly analysed by Actor \cite{Actor},
$A_\mu^{ext}$ does not correspond to a pure gauge. Were it not so,
the introduction  of a chemical potential would not have physical
consequences and this would be the case in any
space-time dimensions. In fact, one cannot gauge away $L_{chem}$
by means of a bounded gauge
transformation. As explained in \cite{hf}, the chiral rotation
which decouples the chemical potential,
although unbounded  can be properly handled by
putting the system in a spatial
box, then introducing adequate counterterms and
finally taking the infinite volume limit.

After the decoupling transformation, the partition function,
can be written in the form
\begin{equation}
Z = {\cal N}\sum_n
\int \D \bar\chi \D \chi \D\phi ~ \exp (- S_{eff}^{(n)})
\label{ef}
\end{equation}
where $ S_{eff}^{(n)}$ is the effective action in the $n^{th}$
topological sector,

\begin{eqnarray}
& & S_{eff}^{(n)}  = \int d^2x ~ \bar\chi (i\!\!\not\!\partial +
{\not\!\! A}^{(n)})\chi -
\frac{N_f}{2e^2}\int d^2x \left(
(\Box\phi)^2 + \epsilon_{\mu\nu}  { F}_{\mu\nu}^{(n)} \Box
\phi \right )   \nonumber \\*[2 mm]
& & -\frac{N_f}{4e^2} \int d^2x {(F_{\mu \nu}^{(n)})}^2
-\frac{tr_f}{2\pi}\int d^2x ~ (\phi\unf+i\M x_1) \Box
(\phi + 2 \phi^{(n)}) + S_c
\label{lar}
\end{eqnarray}

The usual divergency associated to the electromagnetic energy
carried by fermions has to be eliminated by a
counterterm $S_c$ \cite{fks}.
In our approach the divergency manifests through the term
$i\M x_1 \Box \phi^{(n)}$ in eq.(\ref{lar}). Putting the
model in a finite length  box and appropriately adjusting 
$S_c$ yields to a finite answer. 
The counterterm is the Lagrangian counterpart of the one
usually employed in the Hamiltonian approach to handle this problem
\cite{fks}.
In the canonical formulation of QFT this is equivalent to
a redefinition of creation and annihilation operators which
amounts to a shift in the scale used to measure excitations.

As we have mentioned, a fermionic chemical potential ammounts to 
introducing a finite {\em external} charge (i.e. at the spatial 
boundaries) into the theory. In these conditions, 
it can be proved that massless $QED_2$ (and $QCD_2$) at finite density
remains in the Higgs phase. To show this one may compute the string
tension following for instance the procedure described in ref.\cite{Gsm}: 
one starts by integrating out the fermion fields
(or, equivalently, the bosons in the bosonized version) 
in order to derive the effective action for the gauge fields. One
can then compute the Wilson loop to calculate the energy of a couple 
of (static) external charges for a theory containing also dynamical 
`quarks'. 
Now, since zero modes kill the contributions of non-trivial 
topological sectors to  the partition function,
screening can be tested using the effective action in 
the trivial topological sector. In fact, one can see
that for vanishing fermion masses the string tension  vanishes.
In order to discuss these issues  in multiflavour $QED_2$
at finite density, let us note that
after integration of fermions in eq.(\ref{lar}), the 
resulting effective action for
the  gauge field can be written as \cite{RS}
\begin{equation}
S_{eff} = \int d^2x 
(\frac{1}{4e^2} F_{\mu \nu}^2 + \frac{1}{2\pi} a_\mu^2  
- tr_f\frac{i{\cal M}}{2 \pi} \int d^2x\ x^1 F_{01}), 
\label{13}
\end{equation}
where $S_c$ has cancelled the divergent
term, as explained above. Then, at this stage there is no 
divergency to deal with and we can perform our calculation in the whole 
Euclidean space. 
Choosing the Coulomb gauge $a_1 = 0$, appropriate to derive
the static potential between external charges, we obtain from (\ref{13}),
(after integration by parts using zero boundary conditions) the following
effective Lagrangian  
\beq
{\cal L}_{eff} = \frac{1}{2e^2}(\partial_1 a_0)^2 + \frac{1}{2\pi} a_0^2 
- tr_f\frac{i{\cal M}}{2\pi} a_0
\label{15}
\eeq
In order to analize the force between charges, let us pass to
Minkowski space-time, making $a_0 \to i a_0$ so that
the corresponding effective Lagrangian in Minkowski space reads
 \beq
{\cal L}_{M} = -\frac{1}{2e^2}(\partial_1 a_0)^2 - \frac{1}{2\pi} a_0^2 
+ tr_f\frac{{\cal M}}{2\pi} a_0
\label{16}
\eeq
To determine the electrostatic potential between two external charges 
$\pm e'$, we may couple to the gauge field the proper charge 
density
\beq
\rho(x^1) = e'(\delta (x^1+ l) - \delta (x^1 - l))
\label{5}
\eeq
so that the complete effective Lagrangian becomes
\beq
{\cal L} = {\cal L}_M - \rho a_0
\label{6}
\eeq
The resulting equation of motion takes the form
\beq
\partial_1^2 a_0 - \frac{e^2}{\pi} a_0 + \frac{e^2}{2\pi} tr_f\cal{M} =
\rho 
\label{17}
\eeq
and its solution reads
\beq
a_0(x^1) = \frac{e'}{2m} ( \exp(-m \vert x^1+l \vert ) - 
\exp(-m \vert x^1 - l \vert ) ) +tr_f\cal{M}.
\label{18}
\eeq
where $m=e\sqrt{N_f/\pi}$.
The energy of the two test charges at a distance $2 l$
each other, is given by
\beq
V(l)  = \frac{1}{2} \int dx^1 \rho(x^1) a_0(x^1)
\label{9}
\eeq
and we obtain, at finite fermion density, the usual screening potential
\beq
V(l) = \frac{{e'}^2}{2m} (1 - \exp(- 2 m l))
\label{10}
\eeq
with no modification due to the presence of the chemical potential
whose contribution trivially cancels.
We then conclude that in massless multiflavour $QED_2$  
at finite density, any fractional charge $e'$ is screened 
by integer massless charges.  

To get a deeper insight into these results,
let us note that, as it is well-known, only the trivial topological sector
contributes to the partition function for massless fermions
(the contribution of non-trivial sectors being killed by zero-modes).
Then,  deriving the potential between external charges
as we did above or computing the Wilson loop ${\cal W}$ as well, 
one finds that  screening
is not affected by the presence of the chemical potential term
(the Wilson loop calculation yields ${\cal W} = 1$  \cite{Gsm}).
One could argue that, as it happened with the two test
charges, the external charge background associated
to the chemical potential term is itself also screened by massless
dynamical fermions.
After all it is only in the properties of fermion condensates 
that topological sectors enter into play and it is through
its contributions that the chemical potential manifests itself.
One can understand this issue as follows: 
The topological structure of the theory is determined at the boundaries
(recall that in order to calculate the
topological charge one just uses $\oint A^{n}_{\mu} dx_{\mu}=2\pi n$) 
and the corresponding $n$-charge  configurations 
are responsible for the  non-triviallity of the
correlators but not affecting the partition function. It is then precisely
when computing condensates, that the charges at the boundaries
associated with the chemical potential manifest.

To rephrase this analysis,  note that
with the choice of the counterterm  discussed above,
the effective action written in terms of the decoupled fermions
does not depend  on the chemical potentials $\mu_k$. This does not mean
that this term has no physical consequences. In fact,
$\M$ reappears when  computing correlation functions of fermion
fields, once $\bar \psi$ and $\psi$ are written in terms of
the decoupled fields $\bar \chi$ and $\chi$ through eq.(\ref{trafo}).
We shall see in the following sections how fermionic correlators are 
changed, exhibiting oscillatory inhomogeneities in the spatial axes which
depend on $\M$.
The fact that zero modes make certain v.e.v.'s not to vanish,
leads to a highly non-trivial dependence on the chemical potentials.
%%%%%%%%%%%%%%%%%%%%%%%%%%%%%%%%%%%%%%%%%%%%%%%%%%%%%%%%%%%%%%%%%%

\subsection*{(ii) The Correlation Functions}

The introduction of a flavor index implies additional degrees of freedom
which result in $N_f$ independent fermionic field variables. Consequently,
the growing number of Grassman (numeric) differentials calls for additional
Fourier coeficients in the integrand.

It is well known that each coeficient
is related to the quantum numbers of the chosen basis, which is normally
builded up from the eigenfunctions of the Dirac operator. As we have
for one flavor, one has that $n$ of
these eigenfunctions are zero-modes, implying a vanishing
fermionic exponential. Hence, in order to make Grassman integrals
non-trivial, one has to insert several bilinears depending on the
number of zero modes. When the path-integral measure
contains  $N_f$ independent fermionic fields instead of one, the
number of composite insertions is multiplied by  $N_f$ in order to saturate
Grassman integration algebra, with some selection rules which will become
apparent below.

For the sake of brevity let us readily give the result for
general correlation functions of $p$ points with
arbitrary right and left insertions
\beq
C(w_1,w_2,\ldots) = \langle\prod_{k=1}^{N_f}\prod_{i=1}^{r_k}s_+^k(w^i)
\prod_{j=1'}^{s_k}s_-^k(w^j)\rangle
\label{arbab}
\eeq
where
\beq
s_{\pm}^k(w^i)\equiv \bar\psi^k_{\pm}(w^i) \psi^k_{\pm}(w^i) ~,
\eeq
\[ p=\sum_{k=1}^{N_f}p_k \]
and
\[ r_k+s_k=p_k \]
is the total number of insertions in the flavor sector $k$.

After the abelian decoupling, eq.(\ref{arbab}) results in
\begin{eqnarray}
& & C(w_1,w_2,\ldots)  =  \frac{1}{Z}
\sum_{n=1}^{\infty} \! \int \D\phi\
\exp[\frac{N_f}{2\pi}
 \int d^2x\ \phi\Box (\phi+\phi^{(n)})] \times \nonumber\\
& & \exp[-\frac{1}{e^2} \int d^2x\ (\phi+\phi^{(n)})\Box\Box
(\phi+\phi^{(n)})\ ] \times \nonumber\\
& & \exp[ 2\sum_{k=1}^{N_f} (\sum_{i=1}^{r_k}\phi(w^i)
-\sum_{j=1'}^{s_k}\phi(w^j))]  \times \nonumber \\
& &
\prod_{k=1}^{N_f} \exp[ 2 i\mu_k( \sum_{i=1}^{r_k} w^i_1
-\sum_{j=1'}^{s_k} w^j_1)]  \int \D\bar\chi^k \D\chi^k \prod_{i=1}^{r_k}
\bar\chi^k_+(w^i) \chi^k_+(w^i) \times
\nonumber\\
& &
\prod_{j=1'}^{s_k}\bar\chi^k_-(w^j) \chi^k_-(w^j)
\exp[-\int d^2x\ \bar\chi^k (i\ds+\as^{(n)})\chi^k ]
\label{corte}
\end{eqnarray}
where
$w^i_1$ is the space component of $w^i$.
We see from eq.(\ref{corte}) that the chemical potential contribution
is, as expected, completely factorized. Concerning
the bosonic integral, it can be  written as
\begin{eqnarray}
& & B= \!  \exp[{N_f/2\pi \int d^2x\ \phi^{(n)}\Box \phi^{(n)} }] \
\exp[{ -2\sum_{k=1}^{N_f} (\sum_{i=1}^{r_k}\phi^{(n)}(w^i)
-\sum_{j=1'}^{s_k}\phi(w^j)) }] \nonumber \\
& &  \times \exp[{ -2\sum_{k,k'=1}^{N_f} \sum_{i=1}^{p_k}\sum_{i=1}^{p_{k'}}
e_ie_j O^{-1}(w^i,w^j)}]
\label{B}
\end{eqnarray}
with
\[  O^{-1}(w^i,w^j)=K_0(m|w^i-w^j|)+\ln(c|w^i-w^j|). \]
The fermionic path-integral determines the topological sectors
contributing to equation (\ref{arbab}).
More precisely, once the correlator to be computed has
been chosen, Grassman
integration leads to a  non-zero answer only when the number of right
insertions minus the number of left insertions is the same
in every flavor sector. It means that
$r_k-s_k=t\ \ \forall k$, where $t$ is the only topological
flux number surviving the leading sumatory in eq.(\ref{corte}).
(Notice that mixed flavor indices in the elementary bilinear
are avoided, i.e.
we are not including flavor-violating vertices,
in accordance with $QED_4$ interactions).
It is important to stress that each term
explicitly including the classical configuration of the flux sector
cancels out. Consequently, classical configurations
only appear through by means of their
global (topological) properties, namely, through the
difference  in the number of
right and left handed bilinears \cite{bc}.

To conclude, we give the final result for the general correlator
defined in eq.(\ref{arbab}) making use of the explicit form of
Abelian zero modes
\begin{eqnarray}
& & \langle\prod_{k=1}^{N_f}\prod_{i=1}^{r_k}s_+^k(w^i)
\prod_{j=1'}^{s_k}s_-^k(w^j)\rangle=
(-\frac{m e^{\gamma}}{4\pi})^p
\nonumber\\
& &
\exp[ 2 i\sum_{k=1}^{N_f}\mu_k(\sum_{i=1}^{r_k} w^i_1-
\sum_{j=1'}^{s_k} w^j_1)] \prod_{k>k'=1}^{N_f}
\exp [ -4\sum_{i=1}^{p_k}\sum_{j=1}^{p_{k'}} e_i e_j \ln(c|w^i-w^j|)]
\nonumber\\
& &
\exp [ -\sum_{k,k'}^{N_f}\sum_{i=1}^{p_k}\sum_{j=1}^{p_{k'}}
e_i e_j K_0(m|w^i-w^j|)]
\label{arbab2}
\end{eqnarray}
(see Ref.\cite{hf} and \cite{steele} for details).

In order to clearly see the meaning of this expression,
let us show the result for
the simplest non-trivial flavored correlation functions
including mixed right and left handed insertions
\bea
& & \sum_n\langle\bar\psi^1\psi^1(x)\bar\psi^1\psi^1(y)
\bar\psi^1\psi^1(z)
\bar\psi^2\psi^2(w)\rangle{}_n=\nonumber\\
& & 2\cos[\mu_1(z_1-x_1-y_1)-\mu_2 w_1] \langle s_+^1(x)s_+^1(y)
s_-^1(z) s_+^2(w) \rangle{}_1+
\nonumber\\
& & 2\cos[\mu_1(y_1-x_1-z_1)-\mu_2 w_1] \langle s_+^1(x)s_-^1(y)
s_+^1(z)s_+^2(w) \rangle{}_1+
\nonumber\\
& & 2\cos[\mu_1(x_1-z_1-y_1)-\mu_2 w_1] \langle s_-^1(x)s_+^1(y)
s_+^1(z)s_+^2(w) \rangle{}_1,
\eea
\bea
& & \sum_n\langle\bar\psi^1\psi^1(x)\bar\psi^1
\psi^1(y)\bar\psi^2\psi^2(z)
\bar\psi^2\psi^2(w)\rangle{}_n=\nonumber\\
& & 2\cos[\mu_1(x_1-y_1)-\mu_2 (z_1-w_1)] \langle s_+^1(x)s_-^1(y)s_-^2(z)
s_+^2(w) \rangle{}_0+
\nonumber\\
& & 2\cos[\mu_1(x_1-y_1)+\mu_2 (z_1-w_1)] \langle s_+^1(x)s_-^1(y)s_+^2(z)
s_-^2(w) \rangle{}_0+
\nonumber\\
& & 2\cos[\mu_1(x_1+y_1)+\mu_2 (z_1+w_1)] \langle s_+^1(x)s_-^1(y)s_-^2(z)
s_+^2(w) \rangle{}_2
\eea

These expresions make apparent: (i) How the topological
structure of the theory exhibits itself through
the existence of non-trivial  vacuum
expectation values of fermionic bilinears. (Notice that those on the
right hand side are the only surviving terms of the whole sumatory).
(ii) In the multiflavor case, the path-integrals are non-zero
only when the number of right insertions minus the number of
left insertions are identical in every flavor sector.
(iii) The sum over spatial coordinates dramatically exhibits
the translation symmetry breaking discussed above.
(iv) The fixing of various fermion densities
implies a somehow reacher spatial inhomogeneity of the results
with respect to the one flavor case that we have analyzed in \cite{hf},
in the sense that now the ``angles'' depend on various chemical
potentials.
(v) Another difference with respect
to the one flavor case, concerns the trivial cancellation of logarithms 
coming from bosonic and fermionic integration respectively.
Now, this cancellation occuring for one flavor, does not take place
anymore, see eq.(\ref{arbab2}).

%%%%%%%%%%%%%%%%%%%%%%%%%%%%%%%%%%%%%%%%%%%%%%%%%%%%%%%%
\section{Multiflavour ${\em QCD}_2$}

In the present section we consider two dimensional
$SU(N_c)$ Yang-Mills  gauge fields coupled to
massless Dirac fermions in the fundamental representation.
Due to the non-Abelian character of the gauge
symmetry, gluons are charged fields that preserve color flux at each
vertex. Since a colored
quark density is not a quantity to be kept constant,
no chemical potential related to color should be considered but
only that associated with the global symmetry
that yields fermion number conservation. Hence, we
first   include  one chemical potential
term and then  consider a different lagrange multiplier
for each fermionic flavor.

Let us stress that  once the topological effects
arising from vortices are taken into account
and the chemical potential behavior of fermion correlators
is identified, we do not pursue calculations in the bosonic
sector (neither we consider the inclusion of Higgs scalars,
necessary at the classical level for the existence of regular 
vortex solutions).
As we shall see, the boson contribution to the
fermion condensate just factorizes and all the
chemical potential effects can be controlled by
calculations just performed within the fermionic sector.

%%%%%%%%%%%%%%%%%%%%%%%%%%%%%%%%%%%%%%%%%%%%%%%%%%%%%%%%%%%%%%
\subsection*{(i) Handling the Chemical Potential in  $QCD_2$}

We start from the massless ${\em QCD}_2$ (Euclidean) Lagrangian
\beq
L=\bar\psi^{q}(i\partial_{\mu} \gamma_{\mu} \delta^{qq'}+A_{\mu,a}
 t_a^{qq'}\gamma_{\mu}-i\mu\gamma_0\delta^{qq'})\psi^{q'}+
\frac{1}{4g^2} F_{\mu\nu}^a F_{\mu\nu}^a.
\label{lag}
\eeq
where we have included a chemical potential term in the form
\beq
L_{chem}=-i\mu\psi^{\dagger}\psi
\label{lchem}
\eeq
in order to take care of the fermion density constraint.
Here $a=1\dots N_c^2-1,\ $ and $q=1\dots N_c$.
The partition function reads
\beq
Z[\mu] = \int \D \bar\psi \D \psi\D A_\mu \exp[-\int d^2x\, \exp L].
\label{Z}
\eeq
Again, one can decouple the chemical potential
by performing an appropriate chiral rotation for the
fermion variables. Indeed, under the transformation
\begin{eqnarray}
& & \psi = \exp(i\mu \gamma_5 x_1) \chi
\nonumber\\
& & \bar\psi =  \bar\chi\ \exp(i \mu \gamma_5 x_1)
\label{chang}
\end{eqnarray}
 the fermion Lagrangian becomes
\beq
L = \bar \psi \Ds [A, \mu] \psi \to \bar \chi \Ds[A] \chi
\label{trafo}
\eeq
so that the chemical potential completely disappears from the
fermion Lagrangian.
As we have seen, chiral transformations
may generate a Fujikawa jacobian which has to be computed
using some regularization procedure.
For example, using the heat-kernel regularization one introduces
a resolution of the identity of the form
\beq
1 = \lim_{M \to \infty} \exp(- \Ds(\al)^2 /M^2).
\label{yi}
\eeq
where $D_\mu(\alpha)$ ($\alpha \in (0,1)$)
is an interpolating Dirac operator such
that $D_\mu(\alpha = 0) = D_\mu[A, \mu]$ and
$D_\mu(\alpha = 1) = D_\mu[A]$.

After some standard calculation \cite{gmss} one ends with a
Jacobian of the form
\beq
J=\exp \left({i\epsilon_{\mu\nu}/4\pi \int_0^1\, d^2x\ d\al\
tr^c[\mu x_1 F_{\mu\nu}(\al)]}\right)
\label{jac3}
\eeq
where tr$^c$ is the trace with respect to color indices and
\beq
F_{\mu\nu}(\al) = F_{\mu\nu}^a(\al) t^a, \ \ \  a=1,2,\ldots,N_c^2 - 1
\label{FF}
\eeq
Now, the color trace in eq.(\ref{jac3}) vanishes and then
the chiral Jacobian is in fact trivial,
\beq
J = 1
\label{J}
\eeq

We can then write the partition function (\ref{Z}) after the
fermion rotation defined in eq.(\ref{change}) in the form
\beq
Z[\mu] = \int \D A_\mu \D \bar \chi \D \chi \exp(-\int d^2x L)
\label{ZZ}
\eeq
As we have seen in the Abelian case, although $\mu$ is
absent from the r.h.s. of eq.(\ref{ZZ})
one should not conclude that physics is independent of the chemical
potential. For correlation
functions of composite operators which are not chiral invariant,
the chemical potential will reappear when rotating the fermion
variables in the fermionic bilinears. As in the Abelian case,
this happens when computing v.e.v.'s of products  $\bar \psi(x)
\psi(x)$
%%%%%%%%%%%%%%%%%%%%%%%%%%%%%%%%%%%%%%%%%%%%%%%%%%%%%%%%%%%%%%%%%%%%%%

\subsection*{(ii)
Correlation functions in $QCD_2$ with chemical potential}

Our main interest is the computation of fermionic correlators
containing products of local bilinears $\bar \psi \psi(x)$
for which
non-trivial topological gauge field configurations,
and the associated Dirac operator zero-modes, will be
crucial to the obtention of non-vanishing results as explained in
refs.\cite{hf},\cite{bc}-\cite{cmst}.

As in  section 3, we start by writing a gauge field belonging to
the $n^{th}$ topological sector, in the form
\beq
A^a_\mu(x) = A^{a (n)}_\mu(x) + a^{a}_\mu(x)
\label{form}
\eeq
where $A^{a (n)}_{\mu}$ is a fixed classical configuration (as
described in section 2.2)
carrying all the topological charge $n$, and $a_{\mu}^a$, will be
the actual integration variable which belongs to the trivial
sector $n=0$.
Then, we decouple
the $a_{\mu}$ field from the fermions
through an appropriate rotation
(the calculation of the
Fujikawa Jacobian being standard since the decoupling
corresponds to the topologically trivial sector).
Now, it will be convenient to choose the background so that
\beq
A^{a (n)}_+ = 0
\label{ba}
\eeq
In this way, the Dirac operator takes the  form\footnote{
We are using $\gamma_0=\sigma_1$ and $\gamma_1=-\sigma_2.$}
\beq
\Ds[A^{(n)} + a] = \left( \begin{array}{cc} 0 & \partial_+ + a_+ \\
 \partial_- + A^{(n)}_- + a_-	 & 0 \end{array} \right)
\label{vierbein}
\eeq
and we are left with the determinant of this operator once
fermions are integrated out
\beq
Z[\mu] = \sum_n\int \D a_\mu
\exp[\frac{1}{4g^2} F_{\mu\nu}^2[A^{(n)}+a]]
\det \Ds[A^{(n)} + a].
\label{ZZss}
\eeq
As before, we have introduced a sum over different topological sectors.
Now,
we shall factor out the determinant in the classical background so as
to control the zero mode problem. Let us start by introducing
group valued fields to represent $A^{(n)}$ and $a_\mu$
\beq
a_+ = i u^{-1} \partial_+ u
\label{u}
\eeq
\beq
a_- = i d(v \partial_- v^{-1}) d^{-1}
\label{vv}
\eeq
\beq
A^{(n)}_- = i d \partial_- d^{-1}.
\label{a}
\eeq
Consider first the light-cone like  gauge choice
\cite{pol}
\beq
A_- = A^{(n)}_-
\label{Pol}
\eeq
implying
\beq
v = I .
\label{aa}
\eeq
In this gauge the Dirac operator (\ref{vierbein})
reads
\beq
\Ds[A^{(n)} + a]\vert_{lc} =
\left( \begin{array}{cc} 0  & \partial_+ +iu^{-1} \partial_+ u\\
\partial_- + A^{(n)}_- &  0 \end{array} \right)
\label{vier}
\eeq
where subscript $lc$ means that we have used the gauge condition
(\ref{Pol}).
One can easily see (for example by rotating the $+$ sector with
$u^{-1}$ while leaving the $-$ sector unchanged) that
\beq
\det \Ds[A^{(n)} + a]\vert_{lc} =
{\cal N} \det
\Ds[A^{(n)}] \times
\exp(W[u, A^{(n)}]).
\label{pri}
\eeq
Here $W[u,A^{(n)}]$ is the gauged Wess-Zumino-Witten action which
in this case takes the form
\beq
W[u, A^{(n)}] = W[u] + \frac{1}{4\pi}tr_c\int d^2x (u^{-1}
\partial_+ u) (d \partial_- d^{-1})
\label{sa}
\eeq
and $W[u]$ is the Wess-Zumino-Witten action
\beq
W[u] = \frac{1}{2\pi} tr_c
\int d^2 x \partial_{\mu}u^{-1}\partial_{\mu}u+
\frac{e^{ijk}}{4\pi}
tr_c \int_B\! d^3y\,
(u^{-1}\partial_{i}u)(u^{-1}\partial_{j}u) (u^{-1}\partial_{k}u).
\eeq
\label{ww}

Note that in writing the fermion
determinant in the form (\ref{pri}),
the zero-mode problem has been circumscribed
to the classical background fermion determinant.

One can easily  extend the result
(\ref{pri}) to an arbitrary gauge, in terms
of the group-valued fields $u$ and $v$ defined by
eqs.(\ref{u})-(\ref{vv}),
by repeated use of the Polyakov-Wiegmann identity
\cite{polw}
\beq
W[pq] = W[p] + W[q] + \frac{1}{4\pi} tr_c \int d^2x
(p^{-1}\partial_+p)  \, (q \partial_- q^{-1})
\label{PW}
\eeq
The answer is
\beq
\det \Ds[A^{(n)} + a] =
{\cal N} \det \Ds[A^{(n)}] \times \exp(S_{eff}[u,v; A^{(n)}])
\label{sui}
\eeq
\begin{eqnarray}
S_{eff}[u,v; A^{(n)}] & = &
W[u, A^{(n)}]+ W[v] + \frac{1}{4\pi}tr_c\!\int\! d^2x\,
(u^{-1} \partial_+ u) d (v \partial_- v^{-1}) d^{-1}
 \nonumber \\
& & +\frac{1}{4\pi}tr_c\!\int\! d^2x\, (d^{-1} \partial_+ d)
(v \partial_- v^{-1}).
\label{pris}
\end{eqnarray}
Once one has the determinant in the form (\ref{sui}), one can work
with any gauge fixing condition. The gauge choice (\ref{Pol}) is 
in principle not safe since  the corresponding Faddeev-Popov
determinant is $\Delta = \det D_-^{adj}[A^{(n)}]$
implying the possibility of new zero-modes. A more appropriate
choice  would be for example $A_+ = 0$, having
a trivial FP determinant. In any case one ends with a partition 
function showing the following structure
\bea
Z  &=&  \sum_n  \det(\Ds[A^{(n)}])  \int \D a_\mu\,
\Delta\, \delta(F[a])\nonumber\\
& & \exp \left( -S_{eff}[A^{(n)}, a_\mu] - \frac{1}{4g^2}
\int d^2x  F^2_{\mu\nu}[A^{(n)}, a_\mu] \right)
\label{z1}
\eea
Concerning the divergency associated to the external charge distribution,
we have learnt from the Abelian case that one has to carefully handle
this term in order to define excitations with respect to the external
background. In section 3 we have seen that it came from
the interaction of $A^{ext}$ with $F_{\mu\nu}^{(n)}$, appearing in
the fermionic jacobian. Performing a similar calculation in the
present case we would find the non-Abelian analogue of this term
with $tr^c$ acting on it. As we have mentioned above,
this color trace operation implies the vanishing of the corresponding
divergency so that no counterterm might be added in $QCD_2$, meaning that
the relevant vacuum is properly defined.

%%%%%%%%%%%%%%%%%%%%%%%%%%%%%%%%%%%%%%%%%%%%%%%%%%%%%%%%%%%%%%%%

As we have seen, the Lagrangian for $QCD_2$ at finite density
can be written  in terms of $\mu$-rotated fields
which hide the
chemical potential from the partition function. This result however,
does not exhausts the physics of the theory in the sense that
correlation functions {\cal do} depend on $\mu$. Actually, it will
be shown that the chemical potential dependence appears
as a factor  multiplying
the result for correlators of the unconstrained theory.
For this reason,   we shall first describe the computation
of vacuum expectation values of fermion bilinears in the
$\mu = 0$ case and then
consider how this result is modified at finite fermion density.
Hence, we proceed with the analysis of v.e.v's of
products of bilinears like $\bar\chi\chi$. Let us start by
noting that with the choice  (\ref{ba}) for the classical
field configuration, the Dirac equation takes the form
\beq
\Ds[A^{(n)} + a]\left( \begin{array}{c}
			\chi_+ \\
			\chi_-
			\end{array} \right) =
 \left( \begin{array}{cc} 0 & u^{-1}i\partial_+  \\
 dvd^{-1}D_-[A^{(n)}]	 & 0 \end{array} \right)
\left( \begin{array}{c}
			\zeta_+ \\
			\zeta_-
			\end{array} \right)
\label{matrix2}
\eeq
where $\zeta$ is defined as
\bea
& & \chi_+=dvd^{-1}\zeta_+\nonumber\\
& & \chi_-=u^{-1}\zeta_-
\label{lasttrafo}
\eea
so that the Lagrangian in the $n^{th}$ flux sector
can be written as
\bea
L & & =\bar\chi\Ds[a+A^{(n)}]\chi=\zeta_-^*i\partial_+\zeta_- +
\zeta_+^*\Ds_-[A^{(n)}]\zeta_+\nonumber\\
& & \equiv\bar\zeta\ \widetilde D[A^{(n)}]\zeta.
\label{Lzeta}
\eea
In terms of these new fields, the bilinears $\bar\chi\chi$
take the form
\beq
\bar\chi\chi=\zeta_-^*u dvd^{-1}\zeta_+ +\zeta_+^*
dv^{-1}d^{-1}u^{-1}\zeta_- .
\eeq
We observe that the jacobian associated to (\ref{lasttrafo}) 
is nothing else but
the effective action defined in the previous section
by eq.(\ref{pris}).
Hence, an explicit expression for the non-Abelian correlators
reads
\bea
& & \langle \bar\chi\chi(x^1)\dots \bar\chi\chi(x^l)\rangle=
\sum_n\int \D a_{\mu}\ \Delta\ \delta (F[a_{\mu}])\,
\exp[ -S_{eff}(A^{(n)}, a) ]
\nonumber\\
& &
\int \D\bar\zeta \D\zeta\ \exp( \bar\zeta
\left( \begin{array}{cc} 0 & i\partial_+  \\
D_-[A^{(n)}] & 0 \end{array} \right)\zeta )\nonumber\\
& &
B^{q_1p_1}(x^1)\dots B^{q_lp_l}(x^l)\
\zeta_-^{*q_1}\zeta_+^{p_1}(x^1)\dots\zeta_-^{*q_l}\zeta_+^{p_l}(x^l)
+B^{q_1p_1}(x^1)\dots \nonumber\\
& &
B^{-1 q_lp_l}(x^l)\ \zeta_-^{*q_1}\zeta_+^{p_1}(x^1)\dots
\zeta_+^{*q_l}\zeta_-^{p_l}(x^l)
+B^{q_1p_1}(x^1)\dots \nonumber\\
& &
B^{-1 q_{l-1}p_{l-1}}(x^{l-1})B^{-1 q_lp_l}(x^l)\
\zeta_-^{*q_1}\zeta_+^{p_1}(x^1)\dots
\zeta_+^{*q_{l-1}}\zeta_-^{p_{l-1}}(x^{l-1})
\zeta_+^{*q_l}\zeta_-^{p_l}(x^l)
\nonumber\\
& &
+\dots
\eea
where the  group-valued field $B$ is given by
\[ B=u dvd^{-1}. \]

For brevity we have written the gauge field measure in terms 
of the original fields $a_\mu$ although for actual calculations 
in the bosonic sector one has to work using $u$ and $v$ variables 
and proceed to a definite gauge fixing. That is, the measure should 
be written according to
\[ {\cal D}a_{\mu}\rightarrow \D u \D v J_B(u,v,d)\]
and then the gauge condition and Faddeev-Popov determinant
should be included (For example, in the light-cone
gauge $a_+ = 0$, $u = 1$ and the FP determinant is trivial).
Finally, notice that we have obtained a general and completely
decoupled result, from which one sees that due to color degrees
of freedom, the simple product that one
finds in the Abelian case becomes here an involved summatory.

%%%%%%%%%%%%%%%%%%%%%%%%%%%%%%%%%%%%%%%%%%%%%%%%%%%%%%%%%%%%%%

Now that we have an expression for correlators in the unconstrained case, 
let us include the chemical potential in our results. Recall that in
this theory the partition function is (see eq.(\ref{ZZ}))
\beq
Z = \int \D A_\mu \D \bar \chi \D \chi
\exp \left (-\int d^2x\,  \bar\chi (i\ds +\as) \chi
+\frac{1}{4g^2}  F_{\mu\nu} F_{\mu\nu} \right)
\label{ZZs}
\eeq
where $ \bar\chi, \chi$ represent the fermion fields after
the chiral rotation (\ref{chang}) which eliminated the
chemical potential from the Lagrangian.
Since fermionic bilinears can be written as
\[ \bar\psi\psi =\bar\psi_+\psi_+ +\bar\psi_-\psi_-,\]
one has
\beq
\langle\bar\psi\psi\rangle =\exp(2i\mu x_1)\langle\bar\chi_+
\chi_+\rangle +
\exp(-2i\mu x_1)\langle\bar\chi_-\chi_-\rangle.
\eeq

It can be easily seen that the same factorization occurs
when flavor is introduced.
The corresponding transformation for the fermion field isospinor
is now
\begin{eqnarray}
& & \psi = \exp(i\M\unc \gamma_5 x_1) \chi
\nonumber\\
& & \bar\psi =  \bar\chi\ \exp(i \M\unc \gamma_5 x_1)
\label{changes}
\end{eqnarray}
and the bilinear v.e.v takes in this case the form
\beq
\langle\bar\psi\psi\rangle =\exp(2i\M\unc x_1)\langle\bar\chi_+
\chi_+\rangle +
\exp(-2i\M\unc x_1)\langle\bar\chi_-\chi_-\rangle.
\eeq
We shall then include from here on flavor degrees of freedom
with the corresponding constraint on each fermion density.
Since in this case one deals with
$N_f$ fermions coupled to the gauge field,
we can use the fermionic jacobian we have computed for one
flavor to the power $N_f$ while
the bosonic measure remains untouched.
In the light-cone gauge it can be easily seen that
the effective bosonic sector now involves $N_c-1$ massive scalars,
their mass depending on flavor and color numbers by means of a
factor $(2N_c+N_f)^{1/2}$ with respect to the abelian counterpart
(There is also the same number of unphysical massless
particles \cite{lws}).

As we have previously explained, the Dirac operator
has $|n|N_c$ zero modes  in the
$n^{th}$ topological sector, this implying that
more fermion bilinears are needed in order to obtain a non-zero
fermionic path-integral.  Moreover,
since the flavor index implies a factor $N_f$
on the number of Grassman coeficients, the minimal non-zero
product of fermion bilinears
in  the $n^{th}$ sector requires $|n|N_cN_f$ insertions.

Since the properties of the topological
configurations are dictated by those of the torus
of $SU(N_c)$, one can easily extend the results already
obtained for $QED_2$. In particular,
the chirality of the zero modes is dictated by the same index
theorem found in the Abelian theory, this implying that in sector $n>0$
($n<0$) every zero mode has positive (negative) chirality. In this way,
 the right (left) chiral projections of the minimal non-zero
fermionic correlators can be easily computed. One gets
\bea
&  & \langle \prod_k^{N_f}\prod_q^{N_c}\prod_i^{|n|}
\bar\psi^{q,k}_+\psi^{q,k}_+(x^{q,k}_{i})
\rangle{}_n=  \frac{1}{Z^{(0)}}
\!\! \int_{GF} \D u \D v J_B\;
 e^{-S_{Beff}^{(n)}(u,v,d)}\nonumber\\
&  & \prod_k^{N_f}\prod_q^{N_c}\prod_i^{|n|}
\sum_{p_i,l_i}^{N_c}B'{}_k^{q, p_i l_i}(x^{q,k}_{i})\left(
\int \D\bar\zeta
\D\zeta\; e^{\int\bar\zeta\; \!\!\not \widetilde D[A^{(n)}]\zeta}\;
\bar\zeta_+^{p_i}\zeta_+^{l_i} (x^{q}_{i})  \right)_k
\label{gennoab}
\eea
where
\beq
B'{}_k^{q, p_i l_i}(x)=\exp(2i\mu^k x_1)\, u^{p_i q}(x)
(dvd^{-1})^{q l_i}(x),
\eeq
$\bar\zeta_+=\zeta^*_-$ and $\widetilde D[A^{(n)}]$ stands
for the Dirac operator in the r.h.s of eq.(\ref{Lzeta}).
We have used the notation $Z^{(0)}$ for the partition function since
it is completely determined within the $n=0$ sector, see
eq.(\ref{z1}). We have showed every color and flavor indices
explicitly indicating  sum and product operations. The $GF$
label stands for the gauge fixing.
The action 
$S^{(n)}_{Beff}(u,v,d)=N_f S_{WZW}(u,v,d)+S_{Maxwell}(u,v,d)$
is given by the
full gluon field $A^{(n)}(d)+a(u,v)$, and yields a high order
Skyrme-type lagrangian \cite{fns}.

Let us consider $N_c=2$ and $N_f=2$ in order to present the simplest
illustration for the last expression. The minimal fermionic correlator
then  looks
\bea
& & \sum_n\langle\bar\psi^{1,1}_+\psi^{1,1}_+(x^1)
\bar\psi^{1,2}_+\psi^{1,2}_+(x^2)
\bar\psi^{2,1}_+\psi^{2,1}_+(y^1)
\bar\psi^{2,2}_+\psi^{2,2}_+(y^2)\rangle{}_n=\nonumber\\
& &  \frac{1}{Z^{(0)}}\sum_{p,q,r,s}^{N_c=2}
\prod_{k=1}^2\exp[2i\mu^k (x_1+y_1)^k]
\! \int_{GF}\!\!\!\! \D u \D v J_B\; e^{-S_{Beff}^{(1)}(u,v,d)}\,
\times \nonumber\\
& &
B^{1, p q}_k(x^k) B^{2, r s}_k(y^k) \int \D\bar\zeta_k \D\zeta_k\,
e^{\int \bar\zeta_k \widetilde D[A^{(1)}]\zeta_k}\
\bar\zeta_+^{p,k}\zeta_+^{q,k}(x^k)
\ \bar\zeta_+^{r,k}\zeta_+^{s,k}(y^k).
\label{nabex}
\eea

The fermionic path-integral can be easily done, resulting
in the product of eigenfunctions discussed in the sections above,
as follows
\bea
& & \int \D\bar\zeta_k \D\zeta_k\
e^{\int \bar\zeta_k \widetilde D[A^{(1)}]\zeta_k}\
\bar\zeta_+^{p,k}\zeta_+^{q,k}(x^k)
\ \bar\zeta_+^{r,k}\zeta_+^{s,k}(y^k)=
\det\prime(\widetilde D[A^{(1)}])\times\nonumber\\
& & \left( -\bar\eta_+^{(0,1)p,k}\eta_+^{(0,1)q,k}(x^k)
 \bar\eta_+^{(0,2)r,k}\eta_+^{(0,2)s,k}(y^k)
+\bar\eta_+^{(0,1)p,k}\eta_+^{(0,2)q,k}(x^k)\right.
\nonumber\\
& &
\bar\eta_+^{(0,2)r,k}\eta_+^{(0,1)s,k}(y^k)
-\bar\eta_+^{(0,2)p,k}\eta_+^{(0,1)q,k}(x^k)
 \bar\eta_+^{(0,1)r,k}\eta_+^{(0,2)s,k}(y^k)\nonumber\\
& & \left.
+\bar\eta_+^{(0,2)p,k}\eta_+^{(0,2)q,k}(x^k)
 \bar\eta_+^{(0,1)r,k}\eta_+^{(0,1)s,k}(y^k)\right).
\label{ultima}
\eea
Here $\det\prime(\widetilde D[A^{(1)}])$ is the determinat of the
Dirac operator defined in eq.(\ref{Lzeta}) omitting zero-modes and (e.g.)
$\eta^{(0,1)q,k}(x^k)$ is a non-Abelian zero-mode as defined 
in section 2, with an additional flavor index $k$.
Concerning the bosonic sector, the presence of the $F_{\mu\nu}^2$
(Maxwell) term crucially changes the effective dynamics with respect
to that of a pure Wess-Zumino model. One then has to perform
approximate calculations  to compute the bosonic factor,
for example, linearizing the $U$ transformation, see \cite{fns}.
In any case,
once this task is achieved for the $\mu=0$ model,
the modified (finite density) result can be obtained in an exact way.

\section{Summary}

We have presented the correlation functions of
fermion bilinears in multiflavour $QED_2$ and $QCD_2$
at finite fermion density, using a path-integral approach
which is particularly appropriate
to identify the  contributions arising from different topological
sectors. Analysing  correlation functions
for an arbitrary number of fermionic bilinears, we have been
able to determine exactly  its dependence
with the chemical potentials associated to different flavor indices.
As stressed in
the introduction, our work  was
prompted by  recent results by Deryagin,
Grigoriev and Rubakov \cite{dgr}
showing that in the large $N_c$ limit, condensates of $QCD$ in four
dimensions are inhomogeneous and anisotropic at high
fermion density.

Two-dimensional models are a  favorite laboratory to test
phenomena which are expected to happen in $QCD_4$.
In fact, an oscillatory inhomogeneous behavior in
$\langle\bar \psi \psi\rangle$ was found in the Schwinger model
\cite{Kao} using operator bosonization and then the
analysis was completed by finding the exact behavior of
fermion bilinear correlators in \cite{hf}.
Here we have extended this analysis in order
to include flavor and color degrees of freedom within a
path-integral scheme
which makes apparent how topological effects
give rise to the non-triviality of
correlators.

Remarkably,  the oscillatory behavior
related to  the chemical potential that we
have found with no approximation, coincides exactly
with that described in \cite{dgr} for $QCD_4$ within the large
$N_c$ approximation (appart from the anisotropy that of course
cannot be tested in one spatial dimension).
In particular, the
structure of the multipoint correlation functions, given by
eqs.(\ref{arbab2}) and (\ref{gennoab}), shows a non-trivial
dependence on spatial coordinates. This makes apparent that
the ground state has, at finite density, an
involved structure which is a superposition of standing
waves with respect to the order parameter.
Being our model two-dimensional, we were able to control the 
chemical potential matrix behavior in an {\it exact} way so that
we can discard the possibility that the formation of
the standing wave is a byproduct of some approximation. This
should be considered when analysing the results
of ref.\cite{dgr} in $d=4$ dimensions, where one could argue that
use of a ladder approximation as well as the fact
of  neglecting effects subleading in $1/N_c$ 
play an important role in obtaining such a behavior.

Several interesting issues  are open for further investigation
using our approach.
One can in particular study in a very simple way the
behavior of condensates at finite temperature. The chiral anomaly
is independent of temperature and plays a central role
in the behavior of condensates
through its connection with the index theorem.  Therefore,
one should expect that  formulae like (\ref{arbab2}) or (\ref{gennoab})
are valid also for $T > 0$. Of
course, v.e.v.'s at $\mu = 0$ in the r.h.s.
of this equation, should be replaced
by those computed at finite temperature and hence the issue of
zero-modes in a toroidal manifold should be
carefully examined (see e.g. \cite{steele}). At the
light of recent  results concerning $QCD_2$
with adjoint fermions \cite{Gsm,Sm}-\cite{Sm2} it should be of interest
to extend our calculation so as to consider
adjoint multiplets of fermions.

Finally, it should be worthwhile to consider massive fermions and
compute fermion correlation functions at finite density, 
via a perturbative
expansion in the fermion mass following the approach of
\cite{naon}. We hope to report on these problems in a future work.

\section*{Acknowledgements}  The authors would like to thank
Centro Brasileiro de Pesquisas Fisicas of Rio de Janeiro
(CBPF) and CLAF-CNPq,
Brazil, for warm hospitality and financial support.
H.R.C. wish to acknoledge J. Stephany for helpful discussions.
F.A.S. is partially supported
by Fundacion Antorchas, Argentina and a
Commission of the European Communities
contract No. C11*-CT93-0315.

%\end{references}

\end{document}